\documentclass[prl,twocolumn,groupedaddress]{revtex4-1}
\catcode`\@=11
\def\lsim{\mathrel{\mathpalette\@versim<}}
\def\gsim{\mathrel{\mathpalette\@versim>}}
\def\@versim#1#2{\vcenter{\offinterlineskip
\ialign{$\m@th#1\hfil##\hfil$\crcr#2\crcr\sim\crcr } }}
\catcode`\@=12

\catcode`@=12
\usepackage[export]{adjustbox}	
\usepackage{float}
\usepackage{caption}
\usepackage{epsfig,bbm}
\usepackage{amsmath}
\usepackage{slashed}
\usepackage{tikz}
\usepackage{bm}
\usepackage{amssymb}
\usepackage{mathtools}
\usepackage[caption=false]{subfig}
\usepackage{graphicx}
\newcommand{\be}{\begin{equation}}
\newcommand{\ee}{\end{equation}}
\newcommand{\bea}{\begin{eqnarray}}
\newcommand{\eea}{\end{eqnarray}}

\begin{document}
\thispagestyle{empty}
\begin{flushright}
CERN-TH-2019-228
\end{flushright}

\begin{center}
\title{ Chaotic inflation with four-form couplings
}
\author{Hyun Min Lee}
\email[]{hminlee@cau.ac.kr}
\affiliation{Department of Physics, Chung-Ang University, Seoul
06974, Korea \\ CERN, Theory department, 1211 Geneva 23, Switzerland. }
\begin{abstract}
We consider chaotic inflation models with a pseudo-scalar field containing the general couplings to the four-form flux. The four-form mixing with the pseudo-scalar field induces a quadratic potential for inflaton while the coexisting four-form mixing with graviton generates a non-minimal gravity coupling for inflaton. The shift symmetry is respected by the derived inflaton couplings and it is broken spontaneously only by the fixed four-form flux. We discuss the success of inflationary predictions and robustness against higher order terms. Finally, the built-in reheating mechanism is also addressed.  
  
\end{abstract}
\maketitle
\end{center}

\section{Introduction}

Cosmological inflation in the early Universe not only sets up the initial conditions for Big Bang cosmology by solving homogeneity, isotropy, horizon problems, etc, but also provides the seeds for structure formation due to inflaton perturbations. A slowly rolling scalar field, the so called inflaton, is a necessary ingredient, and the shape of the inflaton potential has been constrained by the measurement of anisotropies of Cosmic Microwave Background (CMB) such as Planck \cite{planck2018}.  Some of the previously best known inflation models such as the quartic inflation and the quadratic inflation are now disfavored by CMB measurement \cite{planck2018}.
Typical theoretical problems in chaotic inflation are associated with robustness of the symmetry of inflaton potential and validity of semi-classical description of gravity for trans-Planckian field values.

 Recently, there has been a renewed interest in four-form fluxes as a dynamical relaxation mechanism of cosmological constant and Higgs mass \cite{Dvali,quint,Giudice,Kaloper,hmlee,hmlee2}.   Although the gauge fields corresponding to four-form fluxes are not dynamical in 4D, the four-form fluxes are variable quantities in the process of producing membranes \cite{membrane,tunneling}.  
 Thus, varying four-form fluxes can make cosmological constant or Higgs mass dynamically relaxing to small values as observed. 
 The variable four-form fluxes can also play a role for solving the aforementioned problems in chaotic inflation \cite{inflation}. 
First, the four-form mixing with a pseudo-scalar field induces a quadratic potential for inflaton without an explicit breaking of the shift symmetry. Second, the inflation takes place due to the dominance of the four-form flux in the quadratic potential, allowing for sub-Planckian inflaton field values. Then, the later stage of inflation follows due to the nucleation of membranes reducing the four-form flux. Therefore, the pseudo-scalar with four-form coupling can be a suitable candidate for inflaton while the shift symmetry is spontaneously broken.    

However, a simple four-form coupling to the inflaton induces a quadratic inflaton potential which is not favored by current CMB measurements, so it is necessary to go beyond the quadratic regime. More importantly, in order to fully determine the form of the inflaton potential in chaotic inflation, it is crucial to have quantum gravity corrections beyond the minimal gravity under control. 
In this regard, we note that there are other interesting inflation models leading to inflationary observables that fit the date better than the chaotic inflation types, such as $R^2$ inflation or $\alpha$-attractor models \cite{review}.

Last but not the least, it is important to consider the cosmological evolution of inflaton and Higgs in the context of four-form couplings and set the initial conditions for determining cosmological parameters from structure formation to mass generation in a natural way. For these reasons, it is demanding to investigate general couplings for four-form fluxes beyond the minimal couplings to gravity, maintaining the shift symmetry of the original proposal on chaotic inflation with four-form flux.

 In this article, we consider chaotic inflation models with a pseudo-scalar field coupled to the four-form flux where a non-minimal gravity coupling to the four-form flux plays a crucial role for going beyond the quadratic regime for the inflaton potential.  In essence, the four-form couplings give rise to not only a quadratic potential but also a non-minimal gravity coupling for the pseudo-scalar field, maintaining the shift symmetry.  Then, we show that the shape of the inflaton potential deviates from a quadratic form at large inflaton field values due to the non-minimal gravity coupling.
As a result, we discuss the success of inflationary predictions in the model and testable consequences from the tensor-to-scalar ratio in future CMB experiments. 
Furthermore, we also address safety from unitarity problem and robustness against higher order interactions for inflaton in the inflationary regime where the non-minimal gravity coupling dominates. Finally, we comment on the model-independent mechanism for reheating due to the non-minimal gravity coupling.

 The paper is organized as follows. We first give the model description and derive new interactions for the pseudo-scalar field due to four-form flux. Then, we present new results on the inflationary predictions and discuss unitarity and robustness issues.
 Then, we end up with comments on reheating in this scenario and conclusions are drawn.

\section{The Model}

We consider a pseudo-scalar field $\phi$, a three-index anti-symmetric tensor field $A_{\nu\rho\sigma}$ and its four-form field strength   $F_{\mu\nu\rho\sigma}=4\, \partial_{[\mu} A_{\nu\rho\sigma]}$.
Then, the most general Lagrangian with four-form field couplings to $\phi$  is
\bea
{\cal L} = {\cal L}_0 +{\cal L}_{\rm int}+ {\cal L}_S +{\cal L}_L+ {\cal L}_{\rm memb} \label{full}
\eea
with
\bea
 {\cal L}_0 &=&  \sqrt{-g} \Big[\frac{1}{2}R +\frac{1}{2} \zeta^2 R^2 -\Lambda -\frac{1}{48} F_{\mu\nu\rho\sigma} F^{\mu\nu\rho\sigma}  \nonumber \\
 &&- \frac{1}{2}(\partial_\mu\phi)^2 -V(\phi)\Big], \label{L0} \\
 {\cal L}_{\rm int} &=& \frac{1}{24} \,\epsilon^{\mu\nu\rho\sigma} F_{\mu\nu\rho\sigma} \,(-\alpha R +\mu \phi),  \label{Lagint} \\
 {\cal L}_S &=&\frac{1}{6}\partial_\mu \bigg[\Big( \sqrt{-g}\,  F^{\mu\nu\rho\sigma} + \epsilon^{\mu\nu\rho\sigma} (\alpha R -\mu \phi) \Big)A_{\nu\rho\sigma} \bigg],  \\
 {\cal L}_L &=& \frac{q}{24}\, \epsilon^{\mu\nu\rho\sigma} \Big( F_{\mu\nu\rho\sigma}- 4\, \partial_{[\mu} A_{\nu\rho\sigma]} \Big),  \label{LL} \\
 {\cal L}_{\rm memb}&=& \frac{e}{6} \int d^3\xi\,  \delta^4(x-x(\xi))\, A_{\nu\rho\sigma} \frac{\partial x^\nu}{\partial \xi^a} \frac{\partial x^\rho}{\partial \xi^b} \frac{\partial x^\sigma}{\partial \xi^c} \,\epsilon^{abc}. 
\eea
Here, the scalar potential  $V(\phi)$ could arise due to an explicit breaking of the shift symmetry for the pseudo-scalar field. 
In $ {\cal L}_{\rm int} $ in eq.~(\ref{Lagint}), $\mu$ is a dimensionful four-form coupling to the scalar field  was introduced in the literature \cite{inflation,quint}, and $\alpha$ is a dimensionless non-minimal four-form coupling to gravity \cite{hmlee}.  
${\cal L}_S$ is the surface term for the well-defined variation of the action with the anti-symmetric tensor field, and $q$ in ${\cal L}_L$ (in eq.~(\ref{LL})) is the Lagrange multiplier, and $ {\cal L}_{\rm memb}$ is the membrane action with membrane charge $e$.
Here, $\xi^a$ are the membrane coordinates and $x(\xi)$ are the embedding coordinates in spacetime.
We also note that the $R^2$ term in eq.~(\ref{L0}) is introduced to ensure the stability of the non-minimal four-form coupling to gravity \cite{hmlee}, as will be discussed later.  

Deriving  the equation of motion for $F_{\mu\nu\rho\sigma}$ as follows,
\bea
F^{\mu\nu\rho\sigma}=\frac{1}{\sqrt{-g}}\, \epsilon^{\mu\nu\rho\sigma} \Big( -\alpha R + \mu \phi+q\Big),
\eea
and integrating out $F_{\mu\nu\rho\sigma}$ \cite{inflation}, we obtain the full Lagrangian (\ref{full}) as
\bea
{\cal L} &=&\sqrt{-g} \Big[\frac{1}{2}R+\frac{1}{2} \zeta^2 R^2 -\Lambda 
- \frac{1}{2}(\partial_\mu\phi)^2 -V(\phi) \label{Lagfull} \\
&& -\frac{1}{2} (-\alpha R+ \mu\phi +q)^2  \Big]+ \frac{1}{6}\epsilon^{\mu\nu\rho\sigma} \partial_\mu q A_{\nu\rho\sigma}+ {\cal L}_{\rm memb}.  \nonumber 
\eea 
As a result, the equation of motion for $A_{\nu\rho\sigma}$ makes the four-form flux $q$  dynamical \cite{tunneling}, according to
\bea
\epsilon^{\mu\nu\rho\sigma} \partial_\mu q= -e\int d^3\xi \, \delta^4(x-x(\xi))\, \frac{\partial x^\nu}{\partial \xi^a} \frac{\partial x^\rho}{\partial \xi^b} \frac{\partial x^\sigma}{\partial \xi^c} \epsilon^{abc}. \label{fluxeq}
\eea
The flux parameter $q$ is quantized in units of $e$ as $q=e\,n$ with $n$ being integer. 
Moreover, the four-form flux is reduced after the nucleation of a membrane.

\section{From four-form to non-minimal gravity couplings}

From the result in eq.~(\ref{Lagfull}), we rearrange terms in the following form,
\bea
{\cal L} &=& \sqrt{-g} \bigg[ \frac{1}{2} \Big(  1 + \alpha (\mu \phi+q)\Big) R +\frac{1}{2}(\zeta^2-\alpha^2) R^2  \nonumber \\
&&\quad- \frac{1}{2}(\partial_\mu\phi)^2-V(\phi) -\Lambda-\frac{1}{2} (\mu \phi+q)^2  \bigg].  \label{Lagfull2}
\eea 
The Lagrangian in the above form manifests itself the symmetry structure.
That is, as far as $V(\phi)=0$, there is a shift symmetry for $\phi$ in the full Lagrangian (\ref{Lagfull2}), because the Lagrangian is invariant under $\phi\rightarrow \phi+c$ and $q\rightarrow q-\mu c$ with $c$ being constant. 
But, once $q$ is determined by the equation of motion (\ref{fluxeq}), the shift symmetry is broken spontaneously \cite{inflation}.

Suppose that $\phi$ is stabilized at $\phi_0$ in a false vacuum due to the potential $V(\phi)$ and there are multiple four-form fluxes $q_i$ that are not directly coupled to the inflaton. 
In this case, we can have the effective cosmological constant as $\Lambda_{\rm eff}=\Lambda+V(\phi_0)+\frac{1}{2}(\mu \phi_0+ q)^2+\sum_{i} \frac{1}{2}q^2_i$, so the relaxation of the cosmological constant occurs due to the four-form fluxes \cite{membrane}. The scanning of Higgs mass to a correct value might be also possible in this scenario, thanks to the couplings of multiple four-form fluxes to the Higgs \cite{membrane,tunneling,Giudice,Kaloper,hmlee}. 
After the flux relaxation ends, the inflaton could tunnel from the false vacuum toward the true vacuum, undergoing the slow-roll inflation as we will describe in the following.

Now we further simplify the full Lagrangian (\ref{Lagfull2}) by performing a dual transformation of the $R^2$ term \cite{hmlee}.  In terms of a dual real scalar field $\chi$, then
we obtain the Lagrangian ({\ref{Lagfull2}}) as
\bea
{\cal L} &=& \sqrt{-g} \bigg[ \frac{1}{2}\,\Omega(\phi,\chi,q) R - \frac{1}{2}(\partial_\mu\phi)^2 -V(\phi)  \nonumber \\ 
&&\quad -\Lambda-\frac{1}{2} (\mu \phi+q)^2 -\frac{1}{2} \chi^2 \bigg]  \label{Lagfull3}
\eea
with
\bea
\Omega(\phi,\chi,q)=1 + \alpha \Big(\mu\phi+q\Big)+\sqrt{\zeta^2-\alpha^2}\, \chi. 
\eea
Furthermore, making the field redefinition by
\bea
\sigma= \mu\phi +q+\frac{\sqrt{\zeta^2-\alpha^2}}{\alpha}\, \chi,
\eea
we  rewrite eq.~({\ref{Lagfull3}})  in Jordan frame as
\bea
{\cal L} = \sqrt{-g} \bigg[ \frac{1}{2}\,(1+\alpha \sigma ) R  - \frac{1}{2}(\partial_\mu\phi)^2-V(\phi,\sigma,q)  \bigg] \label{Lagfinal}
\eea
with
\bea
V(\phi,\sigma,q) &=& V(\phi) +\Lambda+\frac{1}{2} (\mu \phi+q)^2 \nonumber \\
&&+\frac{1}{2}\,\frac{\alpha^2}{\zeta^2-\alpha^2} \Big(\sigma-\mu\phi-q \Big)^2. \label{pot}
\eea
As far as $\zeta^2>\alpha^2$, the potential for a new scalar field $\sigma$ is bounded from below, so the stability of the potential is ensured even in the presence of the non-minimal four-form coupling to gravity.  
The sigma field  has a linear non-minimal coupling to gravity, but without a kinetic term. Nonetheless, the sigma field is dynamical due to the kinetic mixing with graviton, as will become clear shortly.

For $\zeta\gtrsim \alpha$ and $\mu\lesssim M_P$, we can integrate out the sigma field, that is, $\sigma=\mu\phi+q$, from the potential for the sigma field in eq.~(\ref{pot}), so $\Omega=1+\alpha(\mu\phi+q)$.  The flux parameter $q$ is fixed by the equation of motion, and we assume $V(\phi)=\Lambda=0$.
Then, making a Weyl scaling of the metric by $g_{\mu\nu}=g^E_{\mu\nu}/\Omega$,
we obtain the effective Lagrangian for a single-field inflation in Einstein frame as follows,
\bea
{\cal L}_E 
=\sqrt{-g_E} \bigg[ \frac{1}{2} R(g_E) -\frac{1}{2}\,K(\phi)(\partial_\mu\phi)^2-V_I(\phi) \bigg] \label{inflatonEFT}
\eea
with
\bea
K(\phi)&=&  \frac{1+\frac{3}{2}\alpha^2\mu^2+\alpha(\mu\phi+q)}{(1+\alpha(\mu\phi+q))^2}, \\
V_I(\phi)&=& \frac{1}{2}\, \frac{(\mu\phi+q)^2}{(1+\alpha(\mu\phi+q))^2}. \label{infpot}
\eea

\begin{figure}
  \begin{center}
    \includegraphics[height=0.35\textwidth]{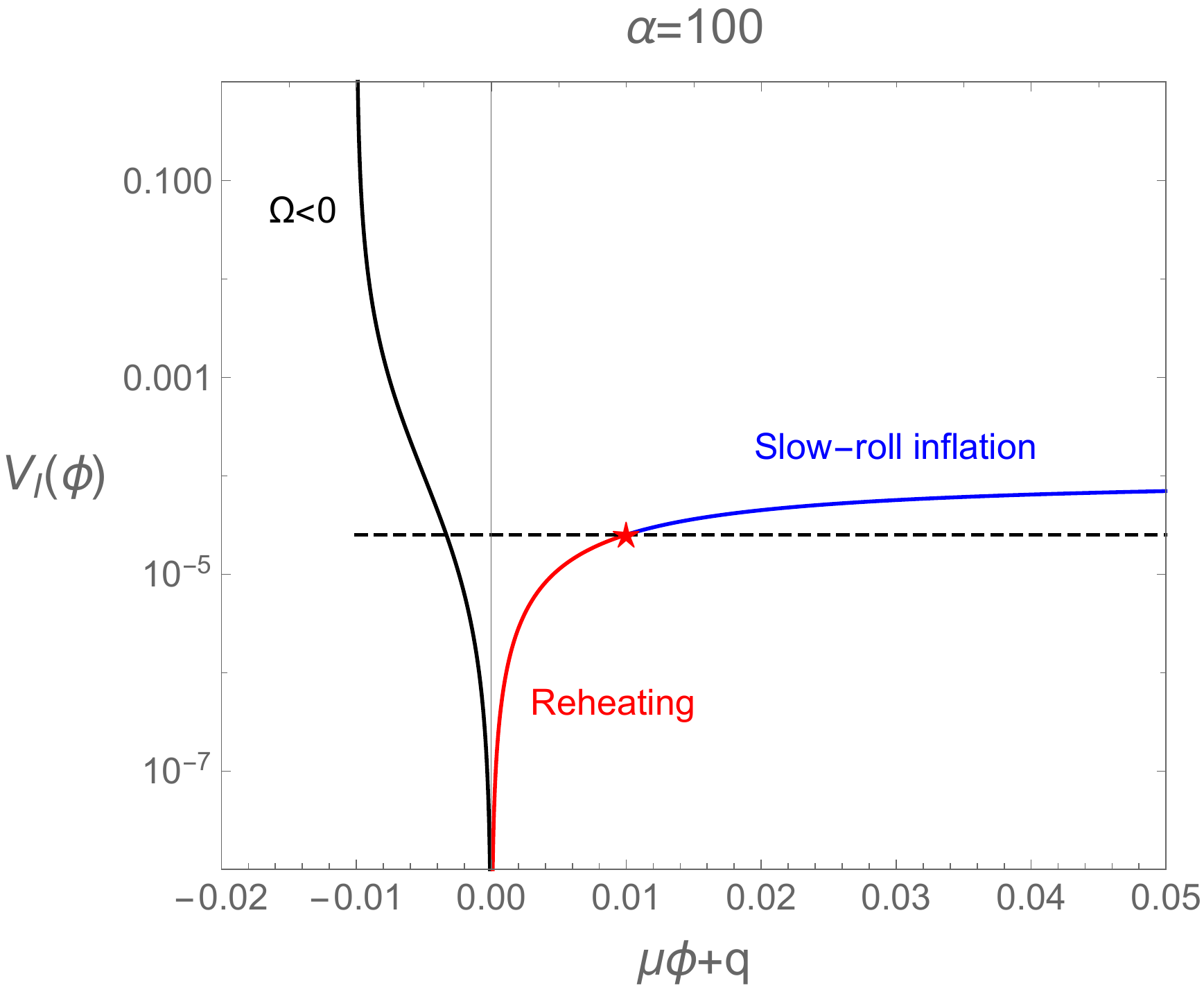}  \vspace{0.5cm} \\
     \includegraphics[height=0.35\textwidth]{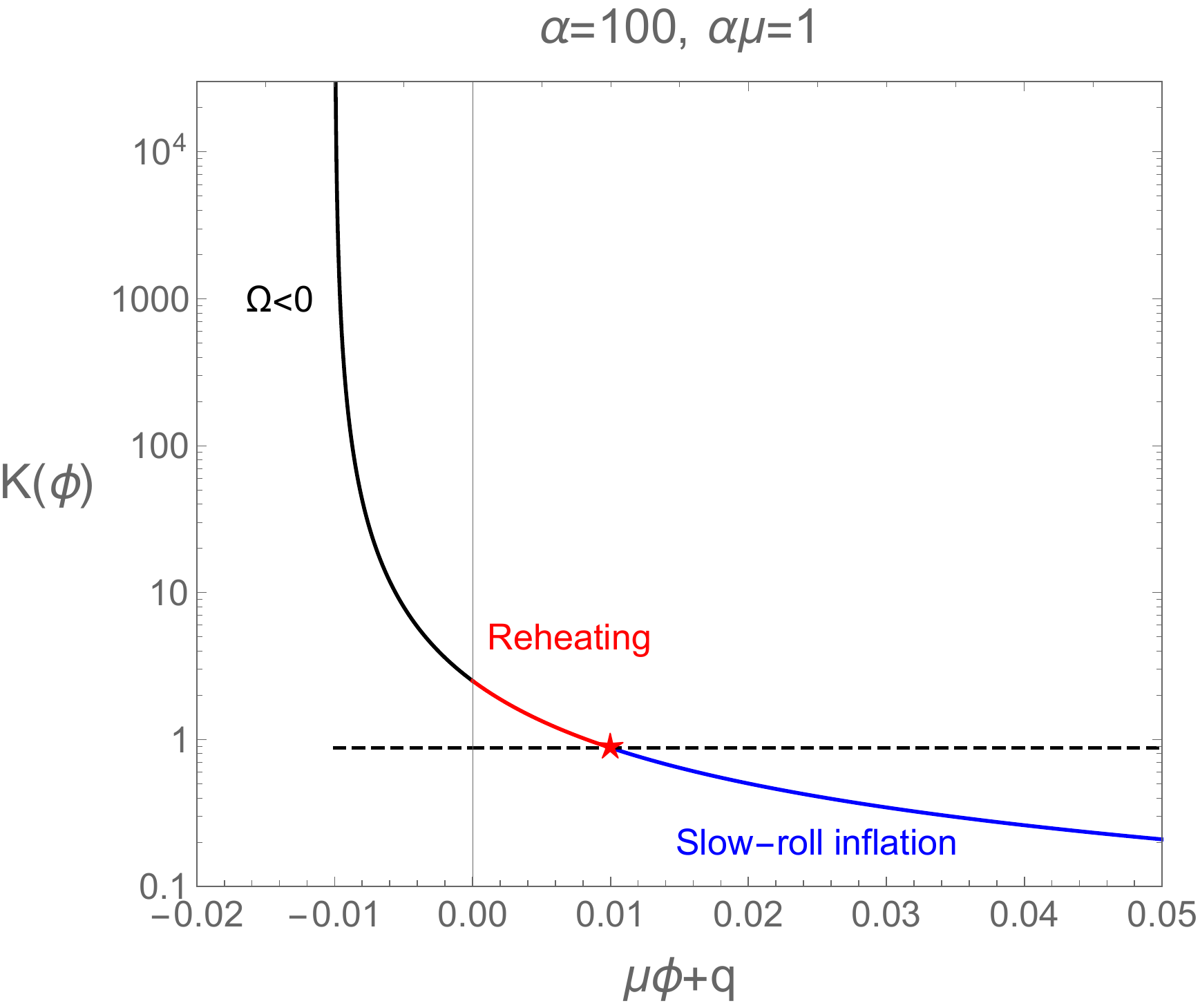}
  \end{center}
  \caption{Upper: Schematic view of the Einstein-frame potential $V_I(\phi)$. The slow-roll inflation takes place along the blue line, inflation ends at the star point, and reheating occurs along the red line. Lower: Schematic view of the non-canonical inflaton kinetic term $K(\phi)$ in Einstein frame, with the same color notations as in the upper plot. In both plots, the region with $\Omega<0$ is not physically viable. }
  \label{potential}
\end{figure}

We note that the frame function $\Omega$ is well-defined only for $\phi>-(q+\alpha^{-1})/\mu\equiv \phi_0$, so the inflaton field space is limited. Thus, during inflation, we restrict ourselves to the field limit with $ \mu\phi+q\gg 1$. In this case, there is no issue in defining the Einstein metric and there is no sudden change of the inflaton direction, so there is no sizable non-Gaussianity in our model.  Furthermore, after inflation ends for $ \mu\phi+q\sim 1$, the inflaton start oscillating near the minimum of the potential, $\phi_{\rm min}=-q/\mu$, but it never crosses the zero of the frame function. Therefore, there is no violent particle production from preheating in our model.  
The schematic view of the inflaton potential and kinetic term in Einstein frame is shown in Fig.~\ref{potential}.

We also remark that the region with $\Omega<0$ as shown in Fig.~\ref{potential} is not physically viable, because the frame function flips the sign to negative, developing the instability problem in gravity.
So, in order to have the well-defined frame function for the entire field space, we can add a quadratic non-minimal coupling of the form, $(\mu\phi+q)^2 R$ \cite{linear,emtrace}.
But, for the later discussion on inflation, we can focus on the region far away from the region with $\Omega<0$, from inflation all the way to reheating. Therefore, we don't introduce such a higher order coupling explicitly in our discussion. But, we will comment on the effects of such a higher order term for inflation dynamics in the later discussion on the robustness of our predictions.

For small inflaton field values and small four-form flux such that $\alpha(\mu\phi+q)\lesssim 1$, we can define the canonically normalized inflaton by ${\bar\phi}=(1+\frac{3}{2}(\alpha\mu)^2)^{1/2}\,\phi$ and identify from eq.~(\ref{inflatonEFT}), higher order terms for the inflaton, $\frac{{\bar\phi}^n}{(\Lambda_n)^{n-4}}$, with $n>4$ and the cutoff scale being given by
\bea
\Lambda_n=M_P \left[\frac{\alpha\mu}{(1+\frac{3}{2}(\alpha\mu)^2)^{1/2}} \right]^{\frac{n}{n-4}}.
\eea
Therefore, for $\alpha \mu\lesssim 1$, the cutoff scale is of order the Planck scale or higher. 
Moreover, interestingly, even for $\alpha\mu\gtrsim 1$, the cutoff scale is saturated to the order of Planck scale independent of $\alpha\mu$, due to a large rescaling of the inflaton kinetic term \cite{rescale,linear0,linear,emtrace}. But, we take $\alpha\mu\lesssim 1$ to keep $\mu\lesssim M_P$ for $\alpha\gtrsim 1$ in the later discussion.

\section{Inflation from four-form couplings}

We are now in a position to discuss the inflationary dynamics in more detail. 

For general inflaton field values, from the Einstein frame Lagrangian in eq.~(\ref{inflatonEFT}), we can define the canonical inflaton field $\varphi$ by
\bea
\varphi&=& 2(\alpha\mu)^{-1} \sqrt{1+\frac{3}{2} \alpha^2\mu^2 +\alpha(\mu\phi+q)} \label{canon} \\
&&-\sqrt{6}\, {\rm arctanh}\bigg[\sqrt{\frac{2}{3}}(\alpha\mu)^{-1}\sqrt{1+\frac{3}{2} \alpha^2\mu^2 +\alpha(\mu\phi+q)}   \bigg]. \nonumber 
\eea
Then, for $u\equiv 1+\alpha(\mu\phi+q)\rightarrow 0^+$, which is the zero of the frame function,  the canonical inflaton becomes $\varphi\approx-\sqrt{\frac{3}{2}}\ln(u)\rightarrow +\infty$, for which the Einstein-frame potential would become $V_I(\varphi)\approx (e^{\sqrt{\frac{2}{3}}\varphi}-1)^2/(2\alpha^2)$, which would diverge.
Therefore, the inflaton should take place far away from the zero of the frame function to support a slow-roll inflation along the blue line in Fig.~\ref{potential} as discussed in the previous section. In this case, even during reheating, the zero of the frame function is never reached. 

If $\alpha(\mu\phi+q)\lesssim 1$ and the four-form dominates the energy density,  the case is the quadratic inflation without a non-minimal four-form coupling to gravity \cite{inflation}, but it is not favored by Planck data \cite{planck2018}.
Thus, we take $\alpha(\mu\phi+q)\gtrsim 1$ with $\alpha\mu\sim 1$ where the non-minimal four-form coupling to gravity becomes crucial. In this limit, we note that only $ \alpha\mu\phi, \alpha q\gtrsim 1$ is required but our following discussion holds, independent of whether the four-form flux is dominant or not. 
In this case, we can find the approximate form of the canonical field $\varphi$ from eq.~(\ref{inflatonEFT}) or from the approximate form of eq.~(\ref{canon}) as
\bea
\mu\phi+q=\frac{1}{4}\alpha \mu^2 \varphi^2.
\eea
Then, the inflaton potential can be now written as
\bea
V_I(\varphi) = \frac{1}{2\alpha^2} \Big(1+\frac{4}{\alpha^2\mu^2\varphi^2} \Big)^{-2}.
\eea
As a result, we obtain the slow-roll parameters as follows,
\bea
\varepsilon &=& \frac{128}{\alpha^4\mu^4 \varphi^6} \,  \Big(1+\frac{4}{\alpha^2\mu^2\varphi^2} \Big)^{-2}, \label{eps} \\
\eta&=& -\frac{48}{\alpha^2\mu^2\varphi^4}  \Big(1-\frac{4}{\alpha^2\mu^2\varphi^2} \Big) \Big(1+\frac{4}{\alpha^2\mu^2\varphi^2} \Big)^{-2}. \label{eta}
\eea
Moreover, the number of efoldings is 
\bea
N&=&\int^{\varphi_*}_{\varphi_f} \frac{d\varphi}{\sqrt{2\varepsilon}} \nonumber \\
&=&\frac{\alpha^2\mu^2}{64} \,\varphi^4_*  \Big(1+\frac{8}{\alpha^2\mu^2\varphi^2_*} \Big) \simeq  \frac{\alpha^2\mu^2}{64}\, \varphi^4_*
\eea
where $\varphi_*$ is the inflation field value at horizon exit and $\varphi_f$ is the inflaton field value at the end of inflation, i.e. $\varepsilon_f=1$ leads to $\varphi_f\approx(128/\alpha^4\mu^4)^{1/6}\simeq 2$ for $\alpha\mu\sim1$.
Therefore, from eqs.~(\ref{eps}) and (\ref{eta}),  we determine the spectral index $n_s$ as
\bea
n_s&=& 1-6\varepsilon_* + 2\eta_* \nonumber \\
&=& 1- \frac{3}{2\alpha\mu}\, \frac{1}{N^{3/2}}-\frac{3}{2N}.
\eea
The tensor-to-scalar ratio is also given by 
\bea
r=16\epsilon_*=\frac{4}{\alpha\mu}\, \frac{1}{N^{3/2}}.
\eea
We note that the measured spectral index and the bound on the
tensor-to-scalar ratio are given by $n_s=0.9670\pm 0.0037$ and $r < 0.07$ at $95\%$ C.L., respectively, from Planck 2018 (TT, TE, EE + low E + lensing + BK14 + BAO) \cite{planck2018}
On the other hand, the normalization of CMB anisotropies with $A_s=\frac{1}{24\pi^2} \frac{V_I}{\epsilon_*}\simeq 2.1\times 10^{-9}$ fixes 
\bea
\alpha=38000 (\alpha\mu)^{1/2} \Big(\frac{N}{50}\Big)^{3/4}.
\eea
Consequently, for $\alpha\mu= 1$ and $N=50(60)$, the spectral index and the tensor-to-scalar ratio become $n_s=0.966(0.972)$ and $r=0.011(0.0086)$, respectively, so the results are in perfect agreement with Planck 2018 within $1\sigma$ \cite{planck2018} and testable in future CMB experiments \cite{future}. In this case, the four-form couplings should be $\alpha=3.8(4.4)\times 10^4$ and $\mu= 6.3(5.5)\times 10^{13}\,{\rm GeV}$.

We comment on the robustness of the inflationary predictions in our model.
Suppose that higher order  terms for the inflaton potential appear only due to the four-form couplings, as follows, 
\bea
\frac{c_n}{\Lambda^{4(n-1)}} \Big( -\frac{1}{24}F_{\mu\nu\rho\sigma} F^{\mu\nu\rho\sigma}\Big)^n
\eea
\bea
= \frac{c_n}{\Lambda^{4(n-1)}}\,\bigg[ (\mu\phi+q)^{2n} - 2n\alpha (\mu\phi+q)^{2n-1}R+\cdots\bigg]\nonumber
\eea 
where $\Lambda$ is the cutoff of the effective theory, $c_n$ is the order one coefficient and the ellipse contains higher curvature terms which are not relevant for the inflation dynamics as we discussed before. As a result, with $\Lambda=M_P$, the inflationary predictions are still valid as far as
\bea
c_n \Big(\frac{M_P}{\sqrt{\alpha}} \Big)^{4(n-1)}\lesssim c_n (\mu\phi+q)^{2(n-1)}\lesssim M_P^{4(n-1)}.
\eea
Here, the lower bound comes from the slow-roll condition and the upper bound comes from the suppression of higher order interactions. Therefore, since $\frac{M_P}{\sqrt{\alpha}}\ll M_P$ for $\alpha\gg 1$ in our case, there is a window of parameter space where the inflationary predictions are robust against higher order interactions. 

Before ending the section, we remark the reheating dynamics in our model. Although the inflaton does not couple directly to the Standard Model in Jordan frame, the inflaton couplings are induced in going to the Einstein frame through the trace of the energy-momentum tensor, $T^\mu_\mu$ \cite{emtrace}, as follows,
\bea
{\cal L}_{\rm int}&=& \frac{1}{2M^2_P}\, \alpha(\mu\phi+q)\, T^{\mu}_\mu,
\eea
which is a Planck-scale suppressed interaction for $\alpha\mu\sim1$. 
During reheating, the inflaton potential becomes quadratic as $V_I\simeq \frac{1}{2}(\mu\phi+q)^2$, so the reheating temperature can be determined from the perturbative decay of the inflation.
Since the inflaton decays dominantly into the SM Higgs and $W, Z$-bosons \cite{emtrace}, the inflaton decay rate is determined to be 
\bea
\Gamma_\phi= \frac{m^3_\phi}{32\pi M^2_P}\, \frac{(\alpha\mu)^2}{1+\frac{3}{2}(\alpha\mu)^2}.
\eea
As a result, for $\alpha\mu=1$, the reheating temperature is given by
\bea
T_{\rm RH}=3.5\times 10^{11}\,{\rm GeV}\, \bigg(\frac{100}{g_*} \bigg)^{1/4} \bigg(\frac{m_\phi}{10^{14}\,{\rm GeV}} \bigg)^{3/2}. \label{RH}
\eea
Therefore, the reheating temperature is large enough to accommodate thermal leptogenesis and thermal production of dark matter. The discussion on those interesting issues will be postponed to a future publication. Finally, we comment that if there is a gluon anomaly interaction for $\phi$ in the form of $(\phi/f_\phi) \, G_{\mu\nu} {\tilde G}^{\mu\nu}$, it could dominate reheating for $f_\phi<M_P$ \cite{inflation}. In this case, reheating temperature could be higher than what we obtained in  eq.~(\ref{RH}), but it is model-dependent. We have just shown the model-independent value for reheating temperature.

\section{Conclusions}

We have suggested a non-minimal four-form coupling to gravity in chaotic inflation scenarios. Incorporating the non-minimal four-form coupling to gravity induces a non-minimal coupling for the pseudo-scalar field respecting the shift symmetry, thus makes the inflaton potential deviate from being quadratic at large field values. In turn, our results are far advanced beyond the regime of the previous consideration without a non-minimal gravity coupling, and the novel outcomes are achieved, such as  the successful inflationary predictions and the testable consequences such as tensor-to-scalar ratio. 

We showed that there is no unitarity violation below the Planck scale in our model. Moreover, we found that higher order terms for the inflaton appear in a particular form respecting the shift symmetry, thus they are well under control  in the inflationary regime where the non-minimal gravity coupling for the four-form flux dominates.
The four-form fluxes can be an important ingredient for the unified description of cosmology, from inflation towards the relaxation of cosmological constant and Higgs mass.

\textit{Acknowledgments}--- 
The work is supported in part by Basic Science Research Program through the National Research Foundation of Korea (NRF) funded by the Ministry of Education, Science and Technology (NRF-2019R1A2C2003738 and NRF-2018R1A4A1025334). 

\end{document}